# Evidence for electromagnetic granularity in polycrystalline Sm1111 iron-pnictides with enhanced phase purity

A. Yamamoto[1], J. Jiang[2], F. Kametani[2], A. Polyanskii[2], E. Hellstrom[2], D. Larbalestier[2], A. Martinelli[3], A. Palenzona[3], M. Tropeano[3], M. Putti[3]

[1] *Department of Applied Chemistry, University of Tokyo*
*7-3-1 Hongo, Bunkyo-ku, Tokyo 113-8656, Japan*
[2] *Applied Superconductivity Center, National High Magnetic Field Laboratory*
*2031 E. Paul Dirac, Dr., Tallahassee, FL 32310, USA*
[3] *University of Genova and CNR-SPIN*
*Via Dodecaneso, 33 - I16146 Genova, Italy*
yamamoto@appchem.t.u-tokyo.ac.jp, putti@fisica.unige.it

We prepared polycrystalline $SmFeAsO_{1-x}F_x$ (Sm1111) bulk samples by sintering and hot isostatic pressing (HIP) in order to study the effects of phase purity and relative density on the intergranular current density. Sintered and HIPped Sm1111 samples are denser with fewer impurity phases, such as SmOF and the grain boundary wetting phase, FeAs. We found quite complex magnetization behavior due to variations of both the inter and intragranular current densities. Removing porosity and reducing second phase content enhanced the intergranular current density, but HIPping reduced $T_c$ and the intragranular current density, due to loss of fluorine and reduction of $T_c$. We believe that the HIPped samples are amongst the purest polycrystalline 1111 samples yet made. However, their intergranular current densities are still small, providing further evidence that polycrystalline pnictides, like polycrystalline cuprates, are intrinsically granular.

**Introduction**

The discovery of superconductivity in LaFeAs(O,F) with critical temperature $T_c$ of 26 K was announced by Hosono *et al.*[1] in February 2008. This event prompted an army of physicists, chemists and material scientists to study these materials and shortly thereafter many other Fe-based superconductors were discovered that are now grouped into 4 families, which are generally referred to as "1111" for REFeAsO, "122" for $AEFe_2As_2$,[2] "111" for LiFeAs[3] and "11" for Fe(Te,Se)[4] where RE denotes rare earth and AE denotes alkali earth. The critical temperatures, in the optimally doped compounds, range from 19 K to 55 K and have very high upper critical fields approaching ~300 T for 1111[5] and ~100 T for 122 doped with potassium[6] or cobalt[7]. These favorable characteristics make it urgent to explore their potential for applications with the hope that practical conductors can be made with these Fe-based superconductors using simpler processes than are needed with high-temperature cuprate superconductors.[8]

Actually these new Fe-based superconductors share several characteristics with cuprates including: layered structures, coexistence of different orderings, occurrence of superconductivity upon doping, short coherence length and unconventional pairing. In cuprates all these features have been shown to be unfavorable for applications. On the other hand, the available results show that both impurities[9,10] and grain boundaries[11,12] are less detrimental to the superconducting properties of these Fe-based superconductors than to the cuprates. Use of Fe-based superconductors in large-scale applications would be greatly enhanced if polycrystalline samples were not intrinsically electromagnetically granular, as is the case for the cuprates. However, investigations of the critical current density ($J_c$) of 1111 polycrystalline samples have all shown significant evidence for granularity and low intergranular $J_c$ values.[11,13,14,15,16,17]

In our previous study, evidence for two distinct scales of current flow was found in polycrystalline Sm- and Nd -1111 samples using magneto-optical imaging (MO) and analysis of the field dependence of the remanent magnetization (RM).[11] A global current density $J_c^{global}$ flowing in the whole sample of order 4000 A/cm$^2$ was determined at 4 K in self-field. This value, which appears to be more than one order of magnitude larger than for early results on randomly-oriented polycrystalline cuprates,[18] is still one of the highest for polycrystalline pnictides reported to date.[19,20,21,22,23,24,25] Granularity has so far limited the properties of pnictide wires.[26,27,28,29] Recently a transport critical current of 3750 A/cm$^2$ at 4.2 K was achieved in $Sr_{0.6}K_{0.4}Fe_2As_2$ wires and tapes using the *ex-situ* powder-in-tube method.[30] This critical current density $J_c$ was decreased by one order of magnitude at 1 T and then remained rather constant with increasing field, suggesting a strong weak-link component of $J_c$ that was suppressed at higher magnetic field. Earlier we had investigated the current-blocking mechanisms in polycrystalline Sm1111 samples by combining low-temperature laser scanning microscopy and scanning electron microscopy observations.[31] This study revealed that many grain-to-grain paths switched off when a magnetic field was applied. It also showed that many grain-boundaries were obstructed by the non-superconducting, normal metal, grain boundary-wetting FeAs phase, as well as a large crack density within and between 1111 grains. Thus the active current cross-section was much less than unity since these defects produce a multiply-connected, current-blocking network.

Here we report more recent efforts to synthesize single-phase polycrystalline Sm1111 samples and to improve the connectivity by employing subsequent sintering and hot isostatic pressing (HIP). A detailed investigation of current flow was performed combining remanent magnetization RM analysis and MO imaging to study the local variation of current density and then using detailed microstructural analysis to understand intergranular current flow in samples synthesized with different processing conditions.

**Sample preparation and characterization:**

As-prepared SmFeAsO$_{0.85}$F$_{0.15}$ (Sm1111) samples were synthesized by solid-state reaction at low pressure from Sm, As, Fe, Fe$_2$O$_3$ and FeF$_2$. First SmAs was synthesized from pure elements in an evacuated, sealed quartz tube at a maximum temperature of 550°C. Sm1111 was then synthesized by mixing SmAs, Fe, Fe$_2$O$_3$ and FeF$_2$ powders in stoichiometric proportions to form SmFeAsO$_{0.85}$F$_{0.15}$ (Sm1111), using uniaxial pressing to make powders into a pellet (~10 mm diameter, 3-4 g), and then heat treating the pellet in an evacuated, sealed quartz tube at 1000°C for 24 h followed by furnace cooling. This as-prepared sample, called A1, was cut into several pieces, each 2 mm thick. Some of these pieces were ground into powder, pressed into a pellet, and then sintered at 1250°C for 24 h in an evacuated, sealed quartz tube. This sample is called S1. Other as-prepared samples were HIPped at 900, 1000 and 1200°C for 10 h under 200 MPa at the Applied Superconductivity Center, National High Magnetic Field Laboratory (called H1, H2 and H3, respectively).

Samples were analyzed by powder X-ray diffraction (XRD) using Cu-$K_\alpha$ radiation. Microstructural observations were performed using a field-emission scanning electron microscope (Carl Zeiss 1540 ESB and XB). Magnetization of the samples was measured by a SQUID magnetometer (Quantum Design: MPMS-XL5s) and a 14 T vibrating sample magnetometer (Oxford). Magneto-optical imaging was done with a 5 μm thick Bi-doped garnet indicator film placed directly onto the sample surface. This allowed us to image the normal field component $B_z$ produced by magnetization currents in the sample induced by solenoidal fields of up to 0.12 T applied perpendicular to the imaged surface.[32]

*X-ray analysis and microstructure*

XRD patterns of samples A1, S1 and H2 are shown in Fig. 1. Lattice parameters were calculated by Rietveld refinement to be $a$ = 0.39301 nm, $c$ = 0.84743 nm for A1, $a$ = 0.39308 nm, $c$ = 0.84732 nm for S1 and $a$ = 0.39310 nm, $c$ = 0.84785 nm for H2. Both $a$ and $c$ show a slight increase after sintering and HIP treatment, which indicates a continuous decrease of fluorine content.[33] The XRD pattern of A1 shows a rather pure phase; however, Rietveld refinement of the pattern shows the presence of additional phases (FeAs, SmOF), which did not exceed 10 volume % of the sample. The SmOF phase slightly increased to 6% in H2 from 4.5% in A1. On the other hand, the FeAs phase was not detectable in H2, although measured as 5.5% in A1 before HIPping. The slight increase in SmOF corresponds to a decrease in fluorine content in the superconducting 1111 phase in H2, in good agreement with the shift of lattice parameters.

Figures 2(a)-(h) show SEM images on polished surfaces of A1 (a, b), S1 (c, d), H1 (e, f) and H2 (g, h). Figure 2 (a) shows that A1, with a relative density of 50-60%, has areas of dense Sm1111 grains interspersed with large pores and also that the wetting FeAs phase is present between Sm1111 grains, even in the dense regions, as the magnified image in Fig. 2 (b) shows. As the processing proceeds from sintering to HIPping, the samples become denser, increasing to 75% in S1 to ~90% in H1, H2 and H3. However, S1 still contains pores, as seen in Fig. 2 (c); they are just smaller than in A1 and they locally disconnect the dense Sm1111 regions. The microstructure becomes very dense after HIPping as shown in Figs. 2 (e) and (g). Here all the dark regions are impurity phases, not pores. As seen in Figs. 2 (b) and (d), sintering changes the grain size. Sm1111 grains have a platelet shape. Their initial grain size in A1 is ~1 μm thick and ~5 μm in diameter. After sintering, the grain size increased to 10-20 μm in diameter. There are cracks along some of the grain boundaries in S1 (Fig. 2(d)) but no evidence of the grain boundary wetting FeAs phase. Grinding A1 into powder and repressing the pellet may have created the large number of small pores in S1, resulting in local cracks between the dense regions. The grain size of the samples H1 and H2 are comparable and similar to that of sample A1, however in the former sample grains are almost equiaxed, whereas in the latter they exhibit the typical platelet shape (leading to a microstructure resembling that observed in A1 and S1 samples). In H1 and H2 (see Figs. 2(e-h)) we did not observe the wetting phase or a continuous network of cracks. The microstructures suggest that sintering and HIPping almost eliminate the grain boundary wetting phase, as well as densifying the sample. These were our goals for our new investigation of the intergranular current in iron-pnictides.

*Magnetization measurements*

The temperature dependence of the susceptibility is shown in Fig. 3. The onset $T_c$ is highest at ~50 K in A1, and thereafter decreases continuously through processing to 45 K in S1, and to 37, 45 and 33 K in H1, H2 and H3, respectively. A1 exhibits an extended transition that may be due to insufficient homogenization and electromagnetic granularity. After sintering (S1) or HIPping at 900°C (H1), the transitions sharpen and H1 displays a relatively sharp single-step transition and nearly full shielding. H3 that was HIPped at 1200°C had a significantly suppressed $T_c$ and a broad transition that are probably due to a decrease in fluorine content in the Sm1111 phase.

The field dependence of the magnetic hysteresis loop widths ($\Delta M$) at 5 K is shown in Fig. 4. The hysteresis loop widths change significantly with the different processing procedures. Compared to the width of A1, the width of the larger-grained S1 quadrupled, the width of H2 doubled, the width of H1 was almost unchanged, and the width of H3 was very small. This degradation of properties for the most processed sample H3 is quite evident. It is interesting to note that all the *M-H* curves (not shown here) presented only a small, almost insignificant positive slope with increasing field. Such a slope is often related to the presence of unreacted $RE_2O_3$ magnetic phases and thus we conclude that the amount of magnetic phases is negligible in our samples and it does not increase after processing.

The magnetization hysteresis convolutes contributions from both intergranular and intragranular current flow. For S1 the increased grain size compared to A1 should make a large contribution to the increased hysteresis under the assumption that intragranular $J_c$ values are much higher than intergranular $J_c$ values. In H2, where the grain size does not increase after HIPping, it is more logical to attribute the larger hysteresis to increased intergranular current flow. Comparing the rather similar hysteresis of A1 and H1, we also need to compare their rather different microstructures in Fig. 2. H1 has a smaller grain size but much less porosity and little intergranular normal FeAs phase, both favorable to intergranular currents. If the critical current density $J_c$ is derived from the hysteresis loop width based on the extended Bean model $J_c = 20 \, \Delta M/a(1-a/3b)$ assuming that current flows uniformly over the whole sample, $J_c$ is of the order of $10^3$-$10^4$ A/cm$^2$ at 5 K for S1, A1, H1 and H2 at self-field. These $J_c$ values are however much higher than the intergranular global $J_c$ values derived from remanent magnetization analysis described below, presumably because the magnetization is dominated by the large contribution of the intragranular currents, as will be discussed below.

To study the origin of these changes of magnetization hysteresis, we performed remanent magnetization analysis. Bulk polycrystalline samples were exposed to many cycles of ever increasing magnetic field $H_a$, followed by removal of the field and measurement of the remanent moment, $m_R$.[11] For a pure, homogeneous sample, we expect flux to penetrate when $H_a/(1-D)$ first exceeds the lower critical field $H_{c1}$, where $D$ is the relevant demagnetizing factor. For weakly coupled polycrystals, flux penetration occurs preferentially at grain boundaries, pores or non-superconducting second phases at lower fields than are needed to penetrate into the grains. Figure 5(a) shows the remanent magnetization as a function of maximum applied field for all samples. A1, the least well sintered and most porous sample, shows a well-separated double step transition indicating two distinctly different scales of current flow. S1 shows penetration only at significantly higher field (~0.1 T); however, there is essentially no lower field transition, a result which indicates that sintering destroyed the intergranular current path, developing a large magnetization due to the enhanced intragranular current possible because of the greatly increased grain size seen in Fig. 2 (d). By contrast, double transitions reappear after the HIPping (H1-H3). After treatment at 900 and 1000°C (H1, H2), the peak positions of the lower field transition are slightly shifted to higher field compared to A1, indicating strengthening of the intergranular current. Assuming that current flows over the whole sample, the intergranular $J_c^{global}$ is given by $J_c^{global} = 2H_{peak1}/w$ where $H_{peak1}$ is the lower field peak in the derivative plot (Fig. 5(b)) and $w$ is the sample size. $J_c^{global}$ values yield ~300, ~0, ~400 and ~240 A/cm$^2$ at 5 K for A1, S1, H1 and H2, respectively. On the other hand, peak splitting was observed for the A1 sample, indicating different length scales of the intergranular current, presumably due to the porous microstructure. The higher field peaks in the derivative plots (Fig. 5(b)) yield intragranular $J_c$ values of $10^6$-$10^7$ A/cm$^2$ at 5 K for A1, S1 and H2. Small, higher field peaks observed in the A1 and H1 samples are perhaps due to small grains of less than ~1 μm, whose magnetic penetration depth cannot be negligible at this scale. For sample H3 treated at 1200°C, both the lower and the higher field peaks were suppressed, consistent with the lower $T_c$, changed lattice parameters indicative of loss of fluorine, and the low-field collapse of the magnetization hysteresis in Fig. 4. The evidence for serious degradation of the superconducting properties in this sample is strong.

Magneto-optical (MO) imaging was performed to directly observe the presence or absence of intergranular global currents in the samples. MO images of A1 showed only very weak MO contrast due to the predominantly local nature of current flow. S1 also showed a granular image indicative only of local intragranular current flow (Fig. 6(b)) that is in good agreement with the absence of the lower field peak in the d$m_R$/d$H_a$ – maximum applied field plot shown in Fig. 5(b). By contrast Fig. 6 (e) shows a well developed roof top pattern, indicating bulk-scale current flow in sample H2. The estimated $J_c^{global}$ = 250 A/cm$^2$ from the magnetic profile in Fig. 6(e) is in good agreement with $J_c^{global}$ obtained by remanent magnetization analysis.

**Discussion**

The thermal processing used in this study modified the microstructure and density of the samples in well defined and systematic ways. Both the sintering and the HIPping densified the microstructure and an increased Sm1111 grain size was observed in S1 after sintering. However, they showed opposite effects on the intergranular current. In S1 there was little wetting phase at the grain boundaries, but it appears that the residual grain boundary cracks are responsible for the absence of global current. On the other hand, although HIPed samples H1 and H2 do not show either grain boundary wetting phase, other non-superconducting phases, or grain boundary cracks, the increase in $J_c^{global}$ is still quite small. We believe that the microstructure of our HIPed samples H1 and H2 are some of the best reported in the literature to date from the view point of phase purity and inter-grain connections, since the 1111 polycrystalline sample reported in the literature contain macroscopic impurity phases and/or wetting phases at grain boundaries.[11,31,34] However, the enhanced phase purity and density was bought at the price of lowered $T_c$ and fluorine loss. Given that the highest $J_c^{global}$ of ~4000 A/cm$^2$ was obtained in an earlier dense sample of Sm1111 with $T_c$ of ~55 K[11] and samples H1 and H2 showed $T_c$ of ~37 K and ~45 K, respectively, the lowered $T_c$ could be a crucial factor affecting $J_c^{global}$. $T_c$ is correlated with doping and thus with carrier density, so changes in doping due for instance to changing the fluorine content, are likely to affect the intergranular current density at grain boundaries in a manner analogous to cuprates. In cuprates the carrier density changes substantially with doping,[35] whereas in pnictides the variation of carrier density is less than 50% as can be evaluated going from undoped to optimally doped compounds.[36] Another, more exotic possibility to explain the low $J_c^{global}$ is that doping can switch between high-$T_c$ nodeless and low-$T_c$ nodal pairings, as suggested by Kuroki *et al.*[37] In this case grain boundaries with reduced $T_c$ could exhibit *d*-wave superconductivity, which strongly suppresses intergranular $J_c$.

Compared to our earlier studies of more polyphase Sm-1111 samples,[11,31,34] the present study has examined a systematic progression from similar polyphase to almost phase-pure samples. However, the small $J_c^{global}$ observed in the randomly oriented HIPped samples (H1 and H2) with much higher purity than A1 appear to support our earlier conclusions that intrinsic granularity occurs in 1111, a result explicitly shown in [001] tilt 122 bicrystal films.[12] The reduction of $T_c$ by loss of fluorine presumably shifts the 1111 compound towards the undoped side even more, again leading to the conclusion that pnictides and cuprates are similar in respect of their sensitivity to disorder at grain boundaries. As in the earlier studies too, there is evidence for some finite intergranular current, but its small magnitude does appear to make it necessary to control grain misorientations by developing high texture to obtain high intergranular $J_c$ in polycrystalline pnictides.

**Conclusions**

In summary we have studied the influence of material processing on the intergranular current density of Sm1111 polycrystalline bulk samples. Microstructural studies showed that after sintering or HIPping the Sm1111 polycrystalline samples have dense microstructures with much less impurity phases such as SmOF and wetting FeAs compared to the as-prepared sample. Magnetic measurements showed that these multiply processed samples have slightly depressed $T_c$ and can significantly enhance magnetic hysteresis loop width compared to the as-prepared sample. However the fact that $T_c$ of the HIPped sample is much lower than the samples with optimal doping indicates that the carrier density was suppressed by sintering and HIPping, possibly enhancing the intrinsic current blocking effect at grain boundaries in sintered and HIPped samples. Nevertheless, even with decreased carrier density, the combined magneto-optical and remanent magnetization analyses showed that HIPping enhanced the intergranular current, whereas ambient pressure sintering significantly decreased it. This suggests that having a very dense microstructure, with few pores, and no wetting phase is also crucial to improve $J_c^{global}$ in optimally doped samples. Our study emphasizes the importance of controlling the doping state too in order to minimize weak-link problems at grain boundaries and to improve the intergranular current density of polycrystalline bulk pnictide samples.

**Acknowledgements**

We are grateful to J. D. Weiss at FSU for experimental help. Work at the NHMFL was supported under NSF Cooperative Agreement DMR-0084173, by the State of Florida and by AFOSR under grant FA9550-06-1-0474. One of the authors (A.Y.) thanks the Japan Society for the Promotion of Science and Global COE program (Chemistry Innovation through Cooperation of Science and Engineering) of MEXT, Japan for financial supports. Work at the University of Genova was partially supported by PRIN2008XWLWF9.

**Figure Captions**

**Fig. 1.** X-ray diffraction patterns for samples A1, S1 and H2. Peaks are identified to be main phase SmFeAs(O,F) (indexed) and impurity SmOF (indicated by #).

**Fig. 2.** Low (a, c, e, g) and high (b, d, f, h) magnification secondary electron microscopy images for A1 (a, b) S1 (c, d), H1 (e, f) and H2 (g, h).

**Fig. 3.** Magnetic susceptibility as a function of temperature after zero-field-cooling and then heating with the application of 1 mT field.

**Fig. 4.** Magnetic hysteresis loop widths ($\Delta M$) as a function of field at 5 K for all the samples.

**Fig. 5.** (a) Remanent magnetization ($m_R$) as a function of maximum applied field ($H_a$) and (b) $d(m_R)/dH_a$ as a function of maximum applied field for all the samples. The data for S1 in (a) have been reduced by a factor of 3 to fit on the plot.

**Fig. 6.** Magneto optical images taken from a polished surface of S1 and H2. (a) Optical microscopy image of S1. (b) MO image under $\mu_0 H_{ex} = 0$ mT at $T = 6$ K for S1 field-cooled (FC) in $\mu_0 H_{ex} = 120$ mT. Granular behaviour indicating very weak or no global current can be seen in image (b). (c) Optical microscopy image of H2. (d, e) MO images of different stages of magnetic flux penetration into H2. H2 was for zero-field cooled to 5.8 K. The remnant images after applying $\mu_0 H_{ex} = 3$ mT (d) and 20 mT (e) then reducing the field to 0 mT. A Meissner strip (d) and a rooftop pattern (e) both indicating clear long range current flow can be clearly seen in images (d) and (e).

**Figure 1**

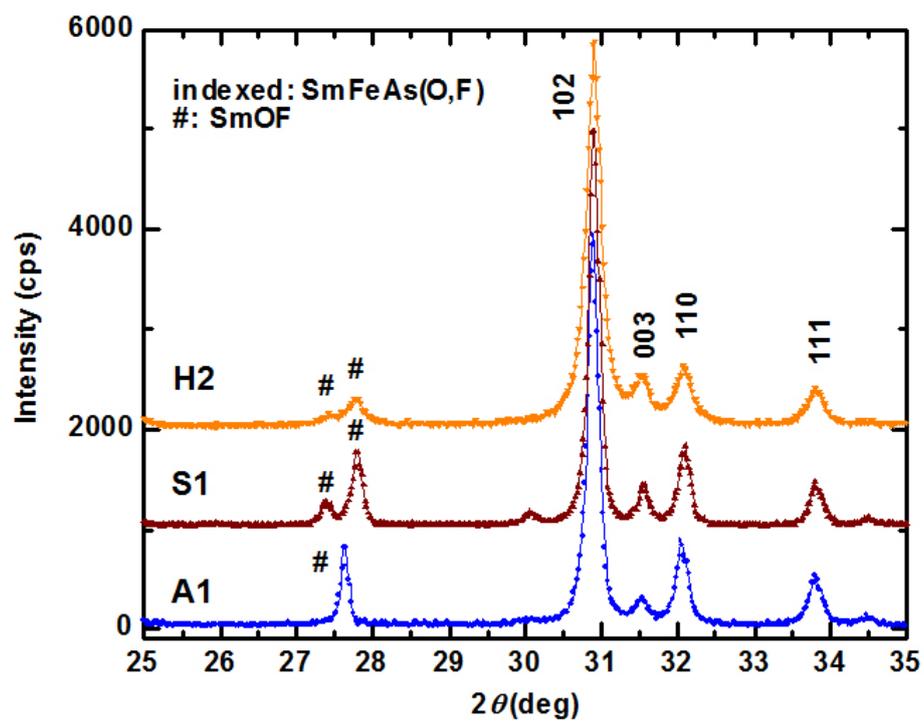

**Figure 2**

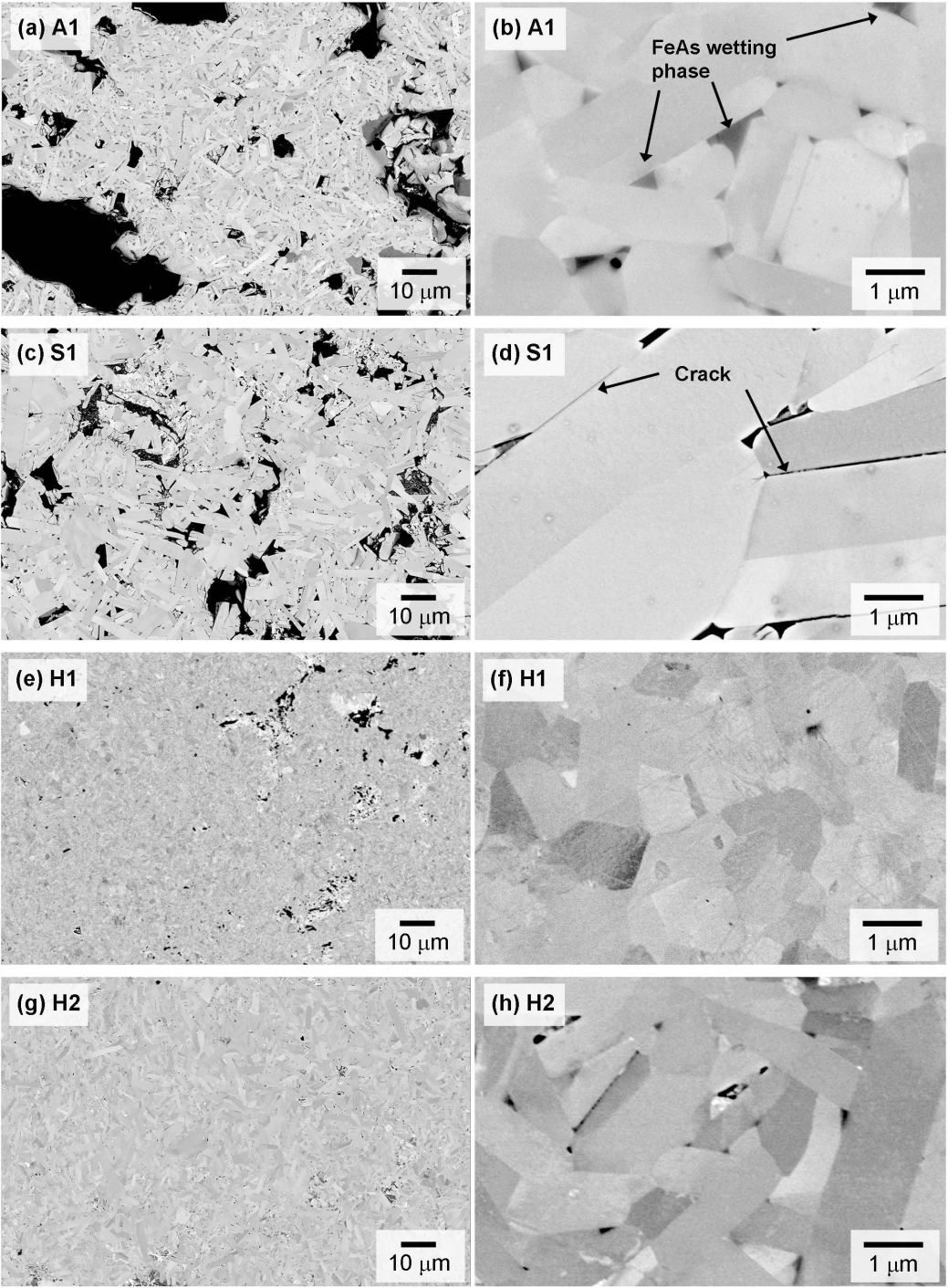

**Figure 3**

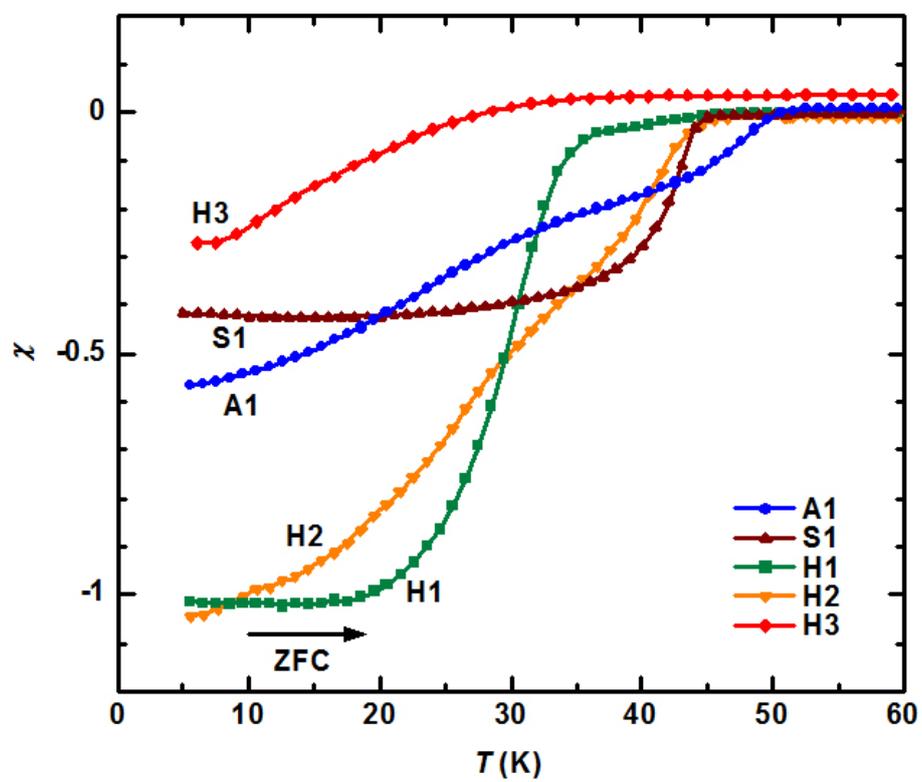

**Figure 4**

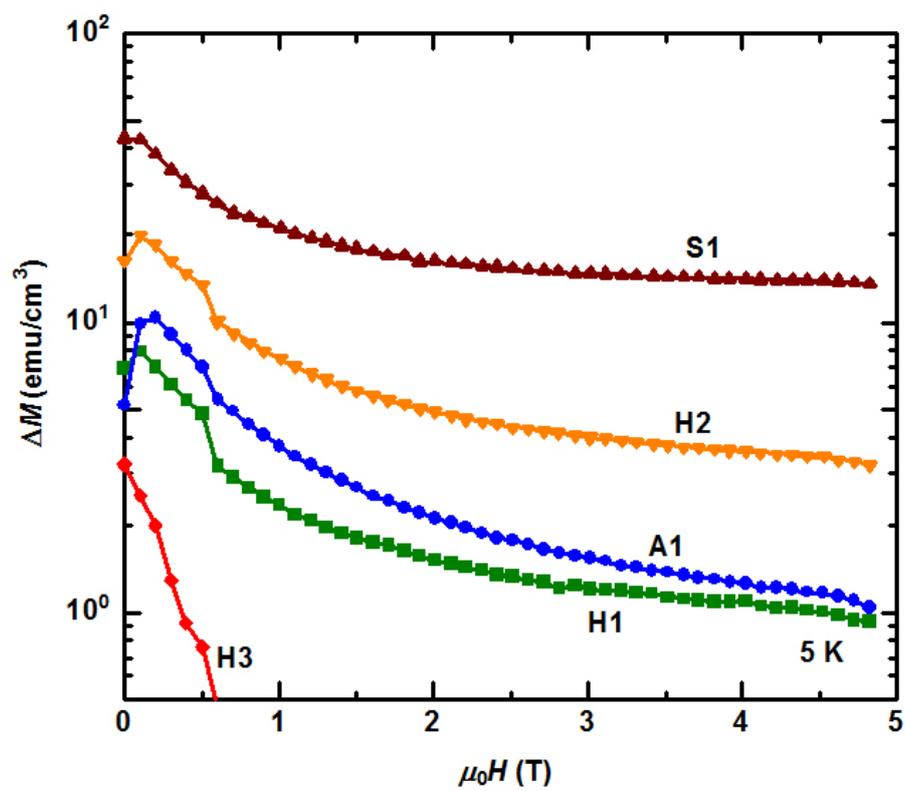

**Figure 5**

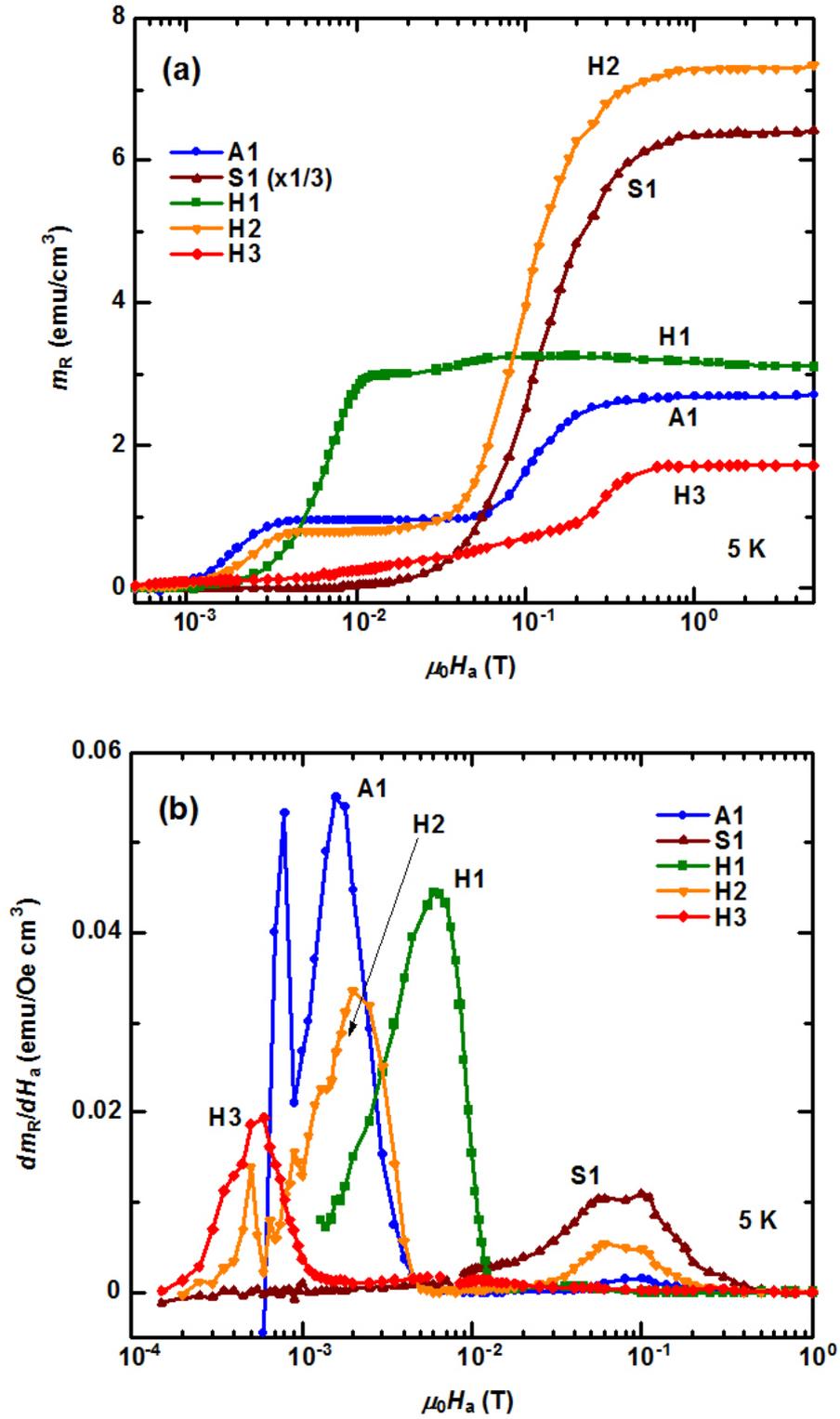

**Figure 6**

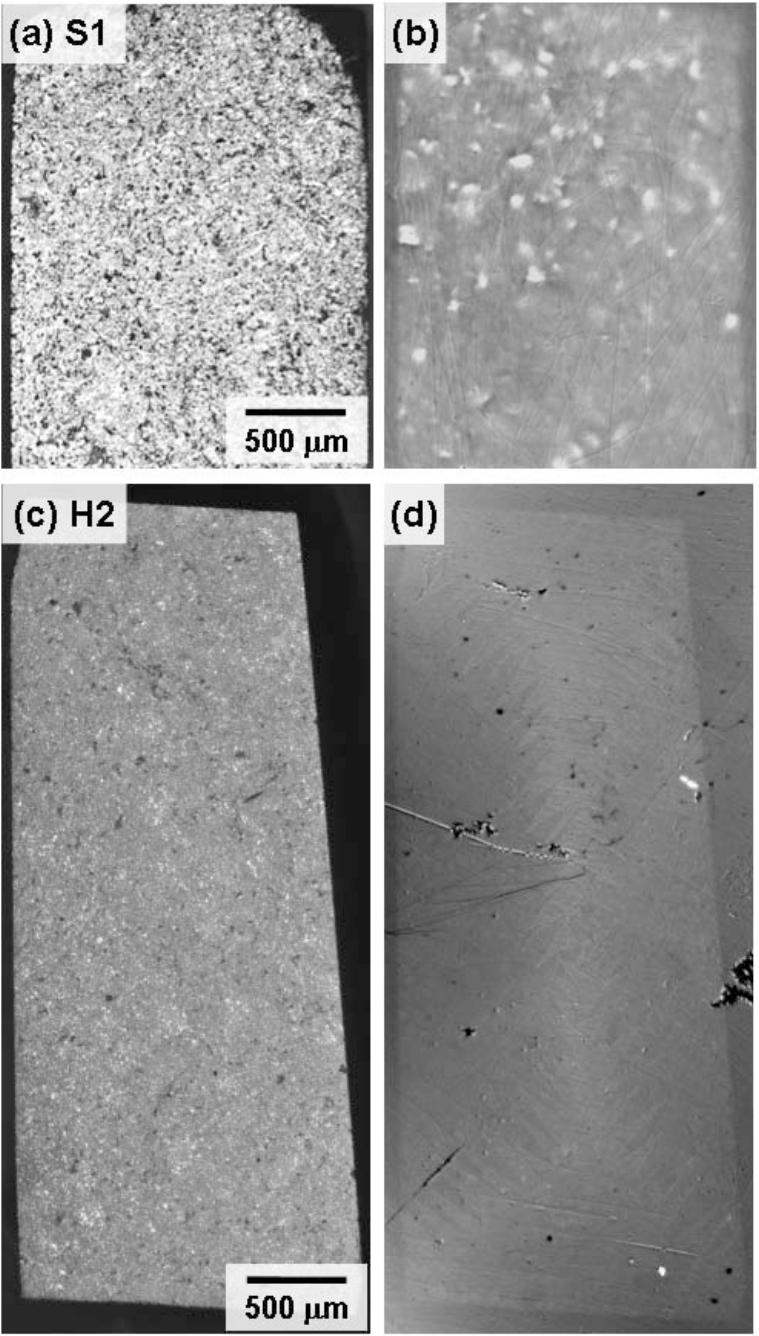